\documentclass[journal]{IEEEtran}

\usepackage{subfigure}

\usepackage{cite}
\ifCLASSINFOpdf
   \usepackage[pdftex]{graphicx}
\else
   \usepackage[dvips]{graphicx}
\fi

\usepackage{amsmath}
\usepackage{algorithmic}

\ifCLASSOPTIONcompsoc
 \usepackage[caption=false,font=normalsize,labelfont=sf,textfont=sf]{subfig}
\else
 \usepackage[caption=false,font=footnotesize]{subfig}
\fi

\usepackage{url}
\usepackage{color}
\usepackage{amssymb}
\usepackage{pifont}
\usepackage{multirow}
\usepackage{arydshln}
\usepackage{mathtools}

\DeclarePairedDelimiterX\Set[1]{\lbrace}{\rbrace}%
 {  #1 }
\newcommand{\ie}{\textit{i.e.~}}
\newcommand{\eg}{\textit{e.g.~}}
\newcommand{\etal}{\textit{et al.~}}

\begin{document}
%
\title{Multi-task pre-training of deep neural networks for digital pathology}
%
%
%

\author{Romain~Mormont, Pierre~Geurts, and Rapha\"el~Mar\'ee}

\IEEEoverridecommandlockouts
\IEEEpubid{\makebox[\columnwidth]{reference coming soon~\copyright2020 IEEE \hfill} \hspace{\columnsep}\makebox[\columnwidth]{ }}

\maketitle

\IEEEpubidadjcol

\begin{abstract}
In this work, we investigate multi-task learning as a way of pre-training models for classification tasks in digital pathology. It is motivated by the fact that many small and medium-size datasets have been released by the community over the years whereas there is no large scale dataset similar to ImageNet in the domain. We first assemble and transform many digital pathology datasets into a pool of 22 classification tasks and almost 900k images. Then, we propose a simple architecture and training scheme for creating a transferable model and a robust evaluation and selection protocol in order to evaluate our method. Depending on the target task, we show that our models used as feature extractors either improve significantly over ImageNet pre-trained models or provide comparable performance. Fine-tuning improves performance over feature extraction and is able to recover the lack of specificity of ImageNet features, as both pre-training sources yield comparable performance.
\end{abstract}


%

\section{Introduction}

\begin{table*}[t]
    \centering
    \caption{Datasets that were used for multi-task pre-training. CLF, DET and SEG respectively stand for \textit{classification}, \textit{detection} and \textit{segmentation}. H\&E, IHC and M3C respectively stand for \textit{hematoxylin and eosin}, \textit{immunohistochemistry} and \textit{Masson's trichrome}. \textit{Images} and \textit{Classes} columns give the number of images and classes of the final (possibly transformed) task for this dataset.} 
    \label{tab:datasets}
    \begin{tabular}{|c|c|l|c|c|c|c|}
        \hline
        \textbf{Name} & Type & \multicolumn{1}{c|}{Task} & Organ \& pathology & Stains & Images & Classes \\
        \hline
MITOS-ATYPIA 14 \cite{roux2014mitos}  & DET &  Detection of mitosis and grading nuclear atypia & breast cancer & H\&E & 64873 & 3 \\
Warwick CRC \cite{sirinukunwattana2016locality}  & DET &  Detection and classification of nuclei & colorectal cancer & H\&E & 2500 & 2 \\
Janowczyk1 \cite{janowczyk2016deep}  & SEG &  Cell nuclei segmentation & breast cancer & H\&E & 31725 & 2 \\
Janowczyk2 \cite{janowczyk2016deep}  & SEG &  Identification of epithelium and stroma & breast cancer & H\&E & 3402 & 2 \\
Janowczyk5 \cite{janowczyk2016deep}  & DET &  Detection of mitosis & breast cancer & H\&E & 24870 & 2 \\
Janowczyk6 \cite{janowczyk2016deep}  & CLF &  Patch classification for WSI segmentation & breast, invasive ductal carci. & H\&E & 277524 & 2 \\
Janowczyk7 \cite{janowczyk2016deep}  & CLF &  Identification of lymphoma subtypes & breast, lymphoma & H\&E & 2244 & 3 \\
Stroma LBP \cite{linder2012identification}  & CLF &  Identification of epithelium and stroma & colorectal cancer & IHC & 2313 & 2 \\
TUPAC2016 Mitosis \cite{veta2019predicting}  & DET &  Detection of mitosis & breast cancer & H\&E & 77853 & 2 \\
BACH18 Micro \cite{aresta2019bach}  & CLF &  Predominant cancer type classification & breast cancer & H\&E & 4800 & 4 \\ 
Camelyon16 \cite{bejnordi2017diagnostic}  & SEG &  Detection of lymph nodes metastases & breast cancer & H\&E & 292226 & 2 \\
UMCM Colorectal \cite{kather2016multi}  & CLF & Tissue type classification & colorectal cancer & H\&E& 5000 & 8 \\
Necrosis \cite{mormont2018comparison}  & CLF &  Necrosed vs healthy tissue & breast cancer & IHC & 882 & 2 \\
ProliferativePattern \cite{mormont2018comparison}  & CLF &  Prolif. vs non-prolif. classification & thyroid cancer & Diff-Quick & 1857 & 2 \\
CellInclusion \cite{mormont2018comparison}  & CLF &  Cell inclusion vs healthy cell classification & thyroid cancer & Diff-Quick & 3637 & 2 \\
MouseLba \cite{mormont2018comparison}  & DET &  Cell classification in bronchoalveolar lavage & lung cancer & MGG & 4284 & 8 \\
HumanLba \cite{mormont2018comparison}  & DET &  Cell classification in bronchoalveolar lavage & lung cancer & MGG & 5420 & 9 \\
Lung \cite{mormont2018comparison}  & CLF &  Tissue subtype classification & lung & H\&E & 6331 & 10 \\
Glomeruli \cite{maree2016approach}  & CLF &  Glomeruli recognition & kidney & M3C & 29213 & 2 \\
Breast1 \cite{mormont2018comparison}  & CLF & Segmentation of cancer tissue & breast cancer & H\&E & 23032 & 2 \\
Breast2 \cite{mormont2018comparison}  & CLF & Segmentation of cancer tissue & breast cancer & H\&E & 17523 & 2 \\
Bone marrow \cite{kainz2017training} & CLF & Cell type classification & bone marrow & H\&E & 1291 & 9 \\
        \hline 
\multicolumn{4}{|c|}{} & \textbf{Total} & 882800 & 81 \\
\hline
    \end{tabular}
\end{table*}

{\let\thefootnote\relax\footnote{Submitted for review on 30 November 2019. Accepted for publication on 1 May 2020. Romain Mormont, Pierre Geurts and Rapha\"el~Mar\'ee are all affiliated with the Department of Electrical Engineering and Computer Science of the University of Li\`ege, 4000, Li\`ege, Belgium. (e-mail: {\tt r.mormont@uliege.be}, {\tt p.geurts@uliege.be} and {\tt raphael.maree@uliege.be}).}}

Recent advances in deep learning have moved forward the field of computer vision. Those developments have been made possible by the availability of large datasets, efficient computing resources and algorithms. These successes have inspired communities to apply deep learning techniques in other fields of application (astronomy, medicine, geography, etc.) where images are prevalent but classical computer vision methods yield unsatisfying performance.

Digital pathology (DP), a domain of medical imaging that focuses on the analysis of large digitized glass slide images (\textit{a.k.a.} whole-slide-images, WSI) containing tissue and cell samples, is no exception. Several groups have applied deep learning on such images for various research and clinical tasks, including cell detection, counting and classification, as well as tissue segmentation. Although deep learning has shown promising results, several difficulties hamper the usefulness of these approaches in practice. One of the main issues is the scarcity of the data \cite{tizhoosh2018artificial,litjens2017survey,robertson2018digital,komura2018machine}. Deep learning algorithms are indeed data-hungry and the number and scale of available datasets is usually much lower in digital pathology than in the natural image domain where these methods have shown the most benefit \cite{deng2009imagenet}. Some of the reasons are the cost and time of the annotation process, which requires the participation of medical experts, but also privacy concerns, which prevent researchers and hospitals from sharing patient data. Moreover, digital pathology images are subject to several sources of variability specific to the process of acquiring samples and turning them into images (tissue preparation protocols including various staining procedures, scanning artefacts, etc.). 

When tackling a new digital pathology classification problem, re-using deep learning techniques developed in other fields does not usually work off-the-shelf because of data scarcity. Therefore, one common approach that has been considered to overcome this issue is transfer learning. The core idea is to pre-train a model on a large dataset (the source task), and then somehow transfer the learned knowledge to facilitate training on a second dataset (the target task). As the source task must be a large dataset, the most common choice is using ImageNet, a classification dataset containing more than 1 million natural images organized into 1000 classes, as a source. Although ImageNet images are very dissimilar to digital pathology images, it has been shown that transferring from the former can still boost performance of deep learning methods on the latter \cite{mormont2018comparison,shang2019and}.

It has also been shown, however, that transfer learning works best when the target task is similar to the source task \cite{yosinski2014transferable}. Whereas task similarity is hard to define formally, it is clear that the ImageNet task is not similar to any digital pathology task. Therefore, this question arises: could we get even better performance from transfer learning by using digital pathology pre-trained models instead of an ImageNet one ? Some works, e.g., \cite{khan2019improving, medela2019few, kraus2017automated, shang2019and}, have advanced that domain-specific pre-training is indeed beneficial but, to the best of our knowledge, there is no in-depth study that attempted to answer this question in digital pathology. The main obstacle preventing this question to be answered is the lack of a large and versatile dataset like ImageNet. However, the digital pathology community has made available many small and medium size datasets through challenges and publications during the last years. This motivates to consider the use of multi-task learning \cite{zhang2017survey} (MTL), a subfield of supervised learning which focuses on methods that solve several tasks simultaneously. Learning several tasks at once presents several advantages such as implicitly regularizing training, therefore helping convergence and reducing overfitting. Provided that enough relevant tasks are available, it makes MTL a great candidate to cope with the data scarcity problem of digital pathology.

Therefore, in this work, we investigate MTL as a way of pre-training neural networks for digital pathology. Our main contributions are as follows. \textbf{(1)} We have collected, assembled and transformed heterogeneous digital pathology datasets into a large versatile pool of classification datasets featuring 22 tasks, 81 classes and almost 900k images (see Section \ref{sec:data}). \textbf{(2)} We have developed a multi-task architecture and a corresponding training scheme for creating a transferable model from these 22 tasks (see Sections \ref{ssec:multitask-architecture} to \ref{ssec:transfer_techniques}). \textbf{(3)} We have developed a robust validation protocol based on a leave-one-task-out scheme for evaluating the transfer performance of our models compared to other approaches (see Sections \ref{ssec:exp:model_selection} to \ref{ssec:exp:parameters}). \textbf{(4)} We have evaluated the performance of the resulting multi-task pre-trained models compared to ImageNet ones, both when pretrained models are used as direct feature extractors and when they are fine-tuned for each target task. We have also compared our approach to a model trained from scratch without any transfer, as well as to a MTL model trained including the target dataset (see Section \ref{sec:results}). \textbf{(5)} Our implementation and multi-task pre-trained models are available on GitHub\footnote{\texttt{https://github.com/waliens/multitask-dipath}}.

\section{Related work}
\label{sec:relatedwork}

Transfer learning is not a recent field of research \cite{pan2010survey} but has grown in popularity with deep learning as it has been shown that features learned on a source task by a neural network could be transferred to a possibly unrelated target task \cite{sermanet2013overfeat, razavian2014cnn, yosinski2014transferable}. The success of this approach is mostly due to the possibility to use the large and versatile ImageNet \cite{deng2009imagenet} dataset as a source task \cite{kornblith2019better}. Medical imaging and digital pathology communities have therefore studied and used transfer learning \cite{tajbakhsh2016convolutional, shin2016deep, mormont2018comparison, babaie2019tissuefold, ponzio2019dealing} as it provides a way of coping with data scarcity. Those works have explored and evaluated different transfer techniques mostly using ImageNet as a source task. The current consensus is that transfer is helpful in most cases and should be considered when tackling a new task. More recently, several works have focused on transferring a model pre-trained on medical, biology or digital pathology datasets. This is motivated by the fact that one can expect better performance from transfer learning when target and source task are close or related \cite{yosinski2014transferable}. In \cite{khan2019improving}, Khan \etal pre-train an InceptionV3 network \cite{szegedy2017inception} on a custom dataset generated from Camelyon16 \cite{bejnordi2017diagnostic} and then transfer the resulting model to a prostate cancer classification task. They show that their pre-trained model outperforms both training from scratch and using an ImageNet pre-trained model. Medela \etal \cite{medela2019few} also make use of transfer learning between two digital pathology tasks but use a different pre-training approach. Indeed, they train a siamese network to distinguish different parts of colorectal tissues. The network is then transferred as a feature extractor on the target task (tumour classification). Shang \etal \cite{shang2019and} use several datasets (including some unrelated to their target task such as \textit{Dogs vs. cats}) and compare ImageNet and domain-specific pre-training in order to tackle colonoscopy image classification. They also show that pre-training on domain-specific data yield superior performance compared to using ImageNet. Kraus \etal \cite{kraus2017automated} train a custom deep neural architecture, DeepLoc, for classifying protein subcellular localization in budding yeast. Then, they assess the transferability of their pre-trained DeepLoc by fine-tuning it on different image sets including unseen classes and show that the pre-training is indeed beneficial.

Independently, multi-task learning \cite{zhang2017survey} has been applied with great successes for a wide-range of application. The success of MTL is notably due to the fact that leveraging several tasks and/or datasets alleviates the need for large amounts of data. Moreover, training in multi-task has regularization effect preventing the model to overfit a particular task therefore yielding a better generalizing model. The modularity of neural networks also allows to embed multi-task specific components hence facilitating its application to deep learning \cite{caruana1997multitask, zhang2017survey}. There are many ways how MTL can be implemented within deep learning with, for instance, architecture tricks \cite{misra2016cross, strezoski2019many} or weight sharing \cite{caruana1997multitask}. 

Multi-task learning has been applied to medical imaging. Samala \etal \cite{samala2017multi} jointly train a classifier on three mammography image datasets (digitized screen-film and digital mammograms) and compare it to single-task training and transfer learning. They show that multi-task trained network generalizes better than the single-task one. Zhang \etal \cite{zhang2016deep} use transfer and multi-task learning to derive image features from Drosophila gene expression. MTL has also been applied more specifically to digital pathology. Pan \etal \cite{pan2018multi} apply MTL for breast cancer classification by using a classification loss and a verification loss. The role of the latter is to ensure that features produced by the network differ for images of different classes. Arvaniti \etal \cite{arvaniti2018coupling} use both weak and strong supervision at once to classify prostate cancer. Shang \etal (mentioned earlier) also evaluate multi-task learning which is the best performing approach on their target task. However, they suggest that more experiments would have to be carried out to assess whether their conclusions are generalizable.

Our work lies at the crossroad of multi-task and transfer learning and differs from the previously presented works mostly on the objective. Indeed, we do not use MTL nor transfer learning for solving a specific task but rather to pre-train a versatile network to be transferred to new tasks. 

\section{Data}
\label{sec:data}

In order to build our pool of tasks, we have collected publicly available datasets (see Table \ref{tab:datasets}) from as many sources as possible. We have also leveraged the Cytomine \cite{maree2016collaborative} platform to collect additional datasets annotated by our collaborators. Some publicly available datasets are missing from our pool because either they could not be converted into a relevant classification problem (e.g. KimiaPath24  \cite{babaie2017classification}, Janowczyk tutorials 3 \& 4 \cite{janowczyk2016deep}) or we could not actually obtain them from the authors (dead link on download page or datasets not released yet \cite{gamper2019pannuke}). Most datasets in our pool are H\&E stained images of human breast cancer but some other organs, pathology and stains are represented, as well as cytology samples, and animal tissues. Also missing in the pool is the BreakHis \cite{spanhol2015dataset} dataset which was kept aside during the development of multi-task training protocol for final model evaluation and comparison to other transfer approaches published using the dataset (see Supplementary Section E). 

For the collected datasets to be used in a multi-task classification setting, some dataset-specific pre-processing procedures had to be executed on most of them. Applying those procedures, we have constructed a pool of 22 classification tasks which contains both binary and multiclass classification problems. The different pre-processing are detailed in Supplementary Section A where selected samples of the final tasks are also provided. Whereas we have tried to avoid intra-dataset class imbalance, there is major inter-dataset imbalance regarding the number of images: the smallest dataset contains 882 images whereas the largest one contains almost 300k. However, we believe it is not an issue and can be made of minor significance by adopting an ad-hoc multi-task training protocol (see Section \ref{ssec:multitask-training}).  
\section{Methods}
\label{sec:methods}

\begin{figure*}
    \centering
    \includegraphics[scale=0.46]{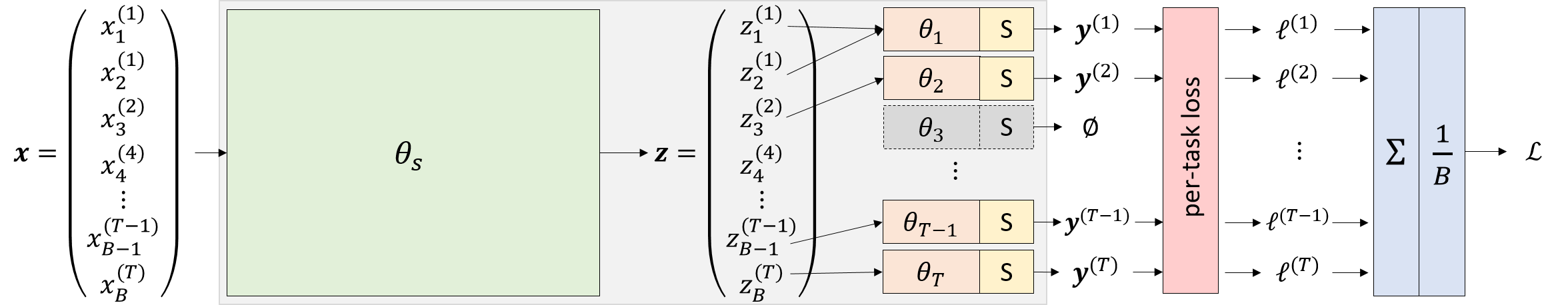}
    \caption{Multi-task architecture. $\mathcal{L}$ is the multi-task loss (see Section \ref{ssec:multitask-training}) and $S$ is a softmax layer. $x_j^{(i)}$ and $z_j^{(i)}$ designate respectively the $j^{\text{th}}$ sample of the batch $\mathbf{x}$ and its corresponding features produced by $\theta_s$. This sample belongs to task $t_i$. Features produced for samples of a given task $t_i$ are routed to this task head $\theta_i$. In this example, there is no sample for task 3 in the batch $\mathbf{x}$. Therefore, the corresponding head $\theta_3$ is inactive (\ie produces no output, no parameters update, no gradient computed) for this iteration.}
    \label{fig:multitask-training}
\end{figure*}

In the following section, we present the training and evaluation protocols and the experiments we have carried out. Those experiments have two main objectives. The first is to evaluate how performance of multi-task and ImageNet pre-trained networks compare when transferred to a target task. The second is to better understand how various training hyperparameters and choices impact the transfer of a multi-task pre-trained network. The multi-task architecture and training are described in \ref{ssec:multitask-architecture} and \ref{ssec:multitask-training}. We present the different transfer techniques we have used in Section \ref{ssec:transfer_techniques}. We have developed a model evaluation and selection protocol which is described in Sections \ref{ssec:exp:transfer_eval} and \ref{ssec:exp:model_eval} whereas the various parameters we have chosen and/or evaluated as well as the experiments we have carried out are presented in Section \ref{ssec:exp:parameters}. 

Regarding notations, we consider a multi-task setting with a pool $\mathcal{P}$ of $T$ classification tasks. Each task $t_i$ has $n_{t_i}$ training samples and its classification problem features $C_{t_i}$ classes. We use $\theta_k$ to designate interchangeably a (part of a) network and its parameters. The notation $|\theta_k|$ designates the number of trainable parameters of the network $\theta_k$. $\mathcal{B}$ represents the set containing all samples from a batch and the batch size is denoted by $B = \left|\mathcal{B}\right|$. $\mathcal{B}_{t_i} \subseteq \mathcal{B}$ is a set that contains all the samples from task $t_i$ in batch $\mathcal{B}$. 

\subsection{Multi-task architecture}
\label{ssec:multitask-architecture}

The structure of our multi-task neural network is similar to those of \cite{shang2019and} and \cite{strezoski2019many} and is guided by the objective of pre-training a network for transfer. Therefore, we have adopted the architecture presented in Figure \ref{fig:multitask-training}. The to-be transferred network is shared for all tasks and denoted by $\theta_s$. We attach a head $\theta_i$ to $\theta_s$ for each task $t_i$ in the pool $\mathcal{P}$. The head $\theta_i$ is simply a fully connected layer of dimensions $f_s\times C_{t_i}$ where $f_s$ is the number of features produced by $\theta_s$. Using such a simple layer has the benefit of making the learning capacity of the heads much lower compared to the shared network (in our experiments $|\theta_s| \gg |\theta_i|, \forall i$), hence forcing $\theta_s$ to learn relevant features for all tasks. In each head, a softmax is attached after the fully connected layer for producing per-task predictions. When forwarding samples in the multi-task network, samples of a given task $t_i$ are only routed through the head $\theta_i$, which outputs predictions for those samples. 

\subsection{Multi-task training}
\label{ssec:multitask-training}

Classical training choices have to be adapted for the multi-task setting. Regarding batch sampling, we have decided to interleave tasks at the sample level, meaning that a single batch can contain samples from different tasks. Indeed, we believe that if batches containing samples from only a single task were alternated, the network would not see a particular task for $T-1$ iterations, with $T$ the number of tasks, which could make the training harder when $T$ is large.

Given the imbalance in terms of number of images per task, a batch sampling procedure had to be carefully established. Indeed, a simple random sampling across all images would prevent the tasks with fewer images from being seen during training. To overcome this issue, we selected each image in a batch by first randomly sampling a task and then sampling an image from this task, thus giving equal weights to all tasks.

Regarding the training loss, we averaged the categorical cross-entropy over all batch samples taking into account their respective tasks. More precisely, the multi-task loss $\mathcal{L}$ is computed from a batch as:
\begin{equation}\label{eqn:loss}
\mathcal{L} = \frac{1}{B} \sum_{i=1}^T \ell^{(i)},\ \ell^{(i)} = - \sum_{j=1}^{\left|\mathcal{B}_{t_i}\right|} \sum_{k=1}^{C_{t_i}} y^{(i)}_{j,k} \log \hat{y}^{(i)}_{j,k}
\end{equation}
where $\ell^{(i)}$ is the loss for the batch samples from task $t_i$.

When developing a multi-task algorithm, a crucial question is what should be shared between tasks. By reducing the capacity of the heads of our architecture, we wanted to enforce the training algorithm to find generic features in $\theta_t$ that work well for all tasks. It might be interesting however to provide a way to slightly relax this implicit constraint, with a hyperparameter. To this end, during training, whereas we train the network with a learning rate $\gamma$, we choose to train the heads with a potentially different learning rate given by $\gamma_h = \gamma \times \tau_\gamma$ where $\tau_\gamma \in \mathbb{R}_{\geq0}$ is a multiplicative factor applied to the global learning rate. This new hyperparameter provides a way of tuning the specificity/genericity of the learned shared features.  Indeed, $\tau_\gamma > 1$ makes the heads learning rate larger and gives therefore more flexibility for the heads to adapt, hence relieving the shared network from learning task-specific features. Taking $\tau_\gamma = 1$ results in using the same learning rate for the whole network.

As previously mentioned, each head $\theta_i$ is randomly initialized, whereas we initialize $\theta_s$ with ImageNet pre-trained weights as it has been shown that doing so accelerates convergence in a single-task setting \cite{mormont2018comparison}. However, it means that trained features of $\theta_s$ are followed directly by the random layers of the heads. This is known to hurt performance in a single-task transfer setting as reported in Yosinski \etal \cite{yosinski2014transferable} and is aggravated in a multi-task setting. Indeed, during the first training iterations, the heads gradients will be relatively large and will work to turn each head weights from random to relevant with respect to its task. However, the resulting back-propagated gradients in the last layer of $\theta_s$ will be an average of the potentially contradictory signals coming from all the heads. In order to attenuate or eliminate this problem, a simple idea consists in making each head weights relevant to its task before training the whole network by running a warm up phase during which $\theta_s$ is frozen (\ie no weights update, no batch normalization update) and only the heads $\theta_i$ are trained with a learning rate $\gamma_w$. 

While preparing our experiments, we have noticed an issue with batch normalization \cite{ioffe2015batch} which is a consequence of the transfer learning settings. Indeed, it has been shown that using a batch normalization-equipped network with datasets across different domains can hurt performance \cite{li2018adaptive, chang2019domain}. We have applied a simple procedure detailed in Supplementary Section B in order to correct the problem. Moreover, the fact that samples are routed in different heads during training requires careful treatment of the gradients after the forward phase. This question is discussed in Supplementary Section C.

\subsection{Transferring a multi-task pre-trained network}
\label{ssec:transfer_techniques}

In this work, we study the two classical approaches of network transfer, namely feature extraction and fine-tuning \cite{litjens2017survey}. In both cases, $\theta_s$ is pre-trained on some source task(s), either ImageNet or several tasks simultaneously in the MTL setting. Feature extraction consists in using the pre-trained $\theta_s$ only to extract the feature vector it outputs for all images of the target task. The extracted features can then be used to learn a third-party classifier, a common choice being a linear SVM \cite{razavian2014cnn, mormont2018comparison}. Fine-tuning consists in further training $\theta_s$ on the target task. A fully connected layer and a softmax are attached to the shared network for generating the target task classes probabilities and the resulting network is trained using for instance stochastic gradient descent. 

Both approaches have their own complementary advantages and drawbacks. As mentioned above, feature extraction uses a linear model which is very fast to train and makes it robust to overfitting when working with small target datasets. However, using fixed pre-trained features makes it possible that the features are not entirely suited for the target task (\eg ImageNet vs. digital pathology), yielding suboptimal performance. Fine-tuning does not suffer from this drawback as the whole network (or a part of it) is retrained on the target task. When using large capacity networks (\eg ResNet or DenseNet), it allows the network to capture and learn task-specific features. This is however an issue when the target dataset is small because the large capacity of the network can lead to overfitting.

\subsection{Evaluating transferability for hyperparameter tuning}
\label{ssec:exp:transfer_eval}\label{ssec:exp:model_selection}

Given a set of tasks available to pre-train a MTL model for future transfer, either by feature extraction or fine-tuning, a question left is how to tune the hyperparameters to train this model. Since we want to optimize the transferability of the model, rather than for example its average performance on the training tasks, we have to design a specific evaluation protocol and a specific metric to assess this transferability for each hyperparameter combination.

For this purpose, we have developed a leave-one-task-out (LOTO) cross-validation scheme, inspired from leave-one-out cross-validation. It consists in removing a set $\mathcal{T} \subset \mathcal{P}$ of one or more tasks from the training pool $\mathcal{P}$, training a multi-task model on $\mathcal{P} \setminus \mathcal{T}$ and then evaluating how the learnt models transfer to the tasks of $\mathcal{T}$. This operation can then be repeated for different $\mathcal{T}$ to increase the stability of the analysis. In our case, we have picked $\mathcal{T}$ to contain only one task when possible. However, it is important that tasks that are closely related are left out together during LOTO cross-validation to avoid leaking shared information between the related tasks during training. In our case, there are two pairs of datasets that are subject to this exception. The first is CellInclusion and ProliferativePattern which are different classification tasks coming from the same WSIs. The second is Breast1 and Breast2 which are the same classification tasks generated with different rules from the same expert annotations. Therefore, applying LOTO exhaustively in our settings leads to 20 possible left out sets $\mathcal{T}$. 

The optimal hyperparameter combination might arguably depend on whether the MTL model will be used for feature extraction or fine-tuning. We have however solely used feature extraction performance as a proxy to evaluate transferability, mainly because we wanted to release a single MTL model for simplicity, but also to reduce the computational costs of our experiments. More precisely, given a left-out set $\mathcal{T}$ and one of its task $t \in \mathcal{T}$, we have evaluated transferability of a multi-task pre-trained network trained on $\mathcal{P} \setminus \mathcal{T}$ by using the resulting $\theta_s$ as a feature extractor on $t$. The training set of $t$ was used to train the features classifier (\ie a linear SVM, see Section \ref{ssec:exp:parameters} for details) and the validation set was used to evaluate it. Transfer performance was evaluated by the  accuracy (ACC) for multi-class classification tasks and the area under the receiver operating characteristic curve (ROC AUC) for binary classification tasks. To cope with the randomness induced by heads initialization and mini-batch sampling, each training of a MTL model was repeated with 5 different random seeds. Note that, at this stage, the test set of $t$ was kept aside for future comparison of a selected multi-task pre-trained network and comparison to ImageNet transfer (see Section \ref{ssec:exp:model_eval}).

All the scores resulting from the same hyperparameters combination but different random seeds can be averaged and the resulting average performance can be used to assess transfer performance on a given task $t$. We need however to aggregate these scores over all (left-out) tasks to assess the overall transferability of a given hyperparameter combination. Averaging ACC and ROC AUC scores over tasks is in general not a good idea as these values depends on the task difficulty and are not directly comparable across tasks. We propose instead to aggregate rankings. More precisely, for each task, the combination of hyperparameters leading to the best model on average was assigned rank $1$ and the worst was assigned the maximum rank. Applying this procedure to all our sets $\mathcal{T}$ produces a rank matrix where $r_{ij}$ is the rank of the $i^{\text{th}}$ combination of hyperparameters evaluated on the left-out task $j$. The best hyperparameter combination is then defined as the one that minimizes the average rank over all left-out tasks:
\begin{equation} \label{eqn:average_ranks}
\overline{r}_i = \dfrac{1}{T_{\text{out}}} \sum^{T_{\text{out}}}_{j = 1} r_{ij}
\end{equation}
where $T_{\text{out}}$ is the number of left-out tasks.

\subsection{Final performance evaluation}
\label{ssec:exp:model_eval}

Our main objective is to compare transfer from multi-task and ImageNet pre-trained networks. In principle, two nested LOTO cross-validation loops should be adopted to carry out such comparison:  for each left-out task in the external LOTO CV loop, an additional internal LOTO CV loop should be run to find the optimal hyperparameter combination for training the MTL model to be transferred to the (external) left-out task. Using two nested loops would be however too expensive computationally\footnote{The simplified scheme we present hereafter has already yielded approximately 20k GPU hours of computation. This computation time would have been multiplied by 10 using two nested LOTO CV loops, which was impossible for us to carry out given our available computing resources.}. We have instead adopted the following simplified scheme. A single LOTO cross-validation is run as described in the previous section. Given a left-out task $t_k$, we select the hyperparameter combination that minimizes the average rank but now excluding task $t_k$ from the average computation (\ie excluding $r_{ik}$ from the calculation of $\overline{r}_i$ in Equation \ref{eqn:average_ranks}). All the models trained using this hyperparameters combination (\ie one per seed) are transferred to the target task $t_k$ using both transfer protocols (\ie feature extraction or fine-tuning). The test set of $t_k$ is used solely to evaluate the resulting transfer performance whereas the training and validation sets can be used by the transfer protocol, feature extraction or fine-tuning, for training and hyperparameter tuning (see Section \ref{ssec:exp:parameters}).

Unlike with a true double LOTO CV loop, the training and validation sets of the left-out task $t_k$ are used, in our simplified scheme, to train some of the MTL models that are transferred to the other tasks for computing the rankings of the hyperparameters combinations for these tasks. However, this is not expected to introduce any bias since the data from task $t_k$ is neither used to decide on the optimal hyperparameter setting of the MTL model transferred to $t_k$ itself (since $r_{ik}$ is excluded from the computation of $\overline{r}_i$), nor to train this MTL model.

\subsection{Hyperparameters settings and experiments}
\label{ssec:exp:parameters} 

\begin{table}
    \centering
    \caption{The multi-task training parameters evaluated using the cross-validation procedure. Using the LOTO scheme, 240 trained models should be transferred to each left out dataset. $\mathcal{H}$ is the set of all hyperparameters combinations.}
    \label{tab:results:parameters}
    \begin{tabular}{|rc|l|c|}
        \hline
        \multicolumn{2}{|c|}{Parameters} & Values & Count \\
        \hline
        Learning rate (LR) & $\gamma$ & $\{10^{-3}, 10^{-4}, 10^{-5}, 10^{-6}\}$ & 4 \\
        LR multiplier & $\tau_\gamma$ & $\{1, 5, 10\}$ & 3 \\
        Shared network & $\theta_s$ & $\{\text{ResNet50}, \text{DenseNet121}\}$ & 2 \\
        Warm up & $w$ & $\{true, false\}$ & 2 \\
        \hline
        \multicolumn{3}{|c|}{Number of combinations $\left|\mathcal{H}\right|$} & 48 \\
        \hline
        \multicolumn{3}{|c|}{with random seeds} & 240 \\
        \hline
    \end{tabular}
\end{table}

The hyperparameters we have studied and their evaluated values are listed in Table \ref{tab:results:parameters}. Parameters values and training choices were established based on early experiments which evaluated multi-task training stability and convergence. Regarding the selected range of learning rates, values higher than $10^{-3}$ resulted in very unstable or diverging trainings whereas values lower than $10^{-6}$ prevented convergence. As shared network $\theta_s$, we have used two popular architectures ResNet50 \cite{he2016deep} and DenseNet121 \cite{huang2017densely}. We have removed the fully connected layer of those networks and replaced it by a global average pooling. Moreover, we have loaded the networks with ImageNet pre-trained weights from PyTorch \cite{paszke2017automatic}.

The multi-task network was trained for 50k iterations using batches of size 64 and SGD as optimizer using momentum set to its default value (\ie $0.9$), learning rate $\gamma$ and heads learning rate multiplier $\tau_\gamma$. Either the whole network was trained directly, or the heads were first warmed up for 5k iterations with learning rate $\gamma_w = 10^{-3}$ before the whole network was trained for 45k iterations.

Classical data augmentation and normalization have been applied to the input images. We have used ImageNet statistics for normalizing the images as early experiments have shown no significant improvement by normalizing with per-task statistics. As data augmentation, we have applied simple random vertical and horizontal flips as well as extraction of a random square crop (if the image is not square already).

When feature extraction was applied either during the evaluation or selection, we have used linear SVM \cite{fan2008liblinear} as feature classifier. Whenever a SVM classifier was trained, we have tuned the $C$ regularization parameter among the following values $\left\{10^{-10}, 10^{-9},...,10^{-1},1\right\}$ by 5-fold cross-validation. The tasks were splitted into folds based on the most relevant information available for the task (patient, slide or image).

Regarding fine-tuning, we have trained the network for 100 epochs on the training set of the target task and have tuned the learning rate among $\{10^{-3}, 10^{-4}, 10^{-5}, 10^{-6}\}$ and have selected the best epoch on the validation set. When transferring from our multi-task pre-trained models, we have performed the selection of the best multi-task models as explained in Section \ref{ssec:exp:model_selection}. Then, for each model trained with the best hyperparameters (\ie one per seed), we have trained one model per fine-tuning hyperparameters combination. For ImageNet, the approach was slightly different as we had only one model to start from. In this case, we have used 5 different seeds for fine-tuning. 

At this point, it is important to note that LOTO cross validation is quite demanding in terms of computing resources as, for all $\mathcal{T}$, all the combinations of hyperparameters and random seeds have to be evaluated. As indicated in Table \ref{tab:results:parameters}, 240 models would have to be trained per set $\mathcal{T}$ of excluded tasks which, given 22 tasks, yields 4800 multi-task trainings. Due to limited availability of computing resources, we had to reduce this number and have done so by reducing the number of left out tasks used in our analysis. In particular, we have kept 10 tasks in 8 sets: $\{\text{CellInclusion}, \text{ProliferativePattern}\}$, $\{\text{Breast1}, \text{Breast2}\}$, $\{\text{MouseLba}\}$, $\{\text{HumanLba}\}$, $\{\text{Necrosis}\}$, $\{\text{Lung}\}$, $\{\text{Glomeruli}\}$ and $\{\text{BoneMarrow}\}$ (see Table \ref{tab:dataset_train_info}). All other tasks were always incorporated however to train each multi-task model.

For the sake of completeness, we have also compared the transfer learning approaches with training from scratch and joint training. The former consists in training a network initialized with random weights. The latter consists in training a network in multi-task using the whole pool of tasks (including the target tasks). 

For training from scratch, we have used the same settings as for the fine-tuning experiment except for the network weights initialization (using initialization strategy as defined in PyTorch): same evaluated learning rates, number of training epochs and same networks. Regarding joint training, we have used the same architecture and training algorithm as for our multi-task pre-training. For the evaluation tasks listed above, only their training set is used for the multi-task training, while their validation and test sets were respectively used for optimizing the hyperparameters and evaluating the selected model. The data for the other training tasks were kept the same as for multi-task pre-training.

\section{Results and discussion}
\label{sec:results}
\begin{figure}[t]
    \center
    \subfigure[DenseNet121]{\includegraphics[scale=0.5]{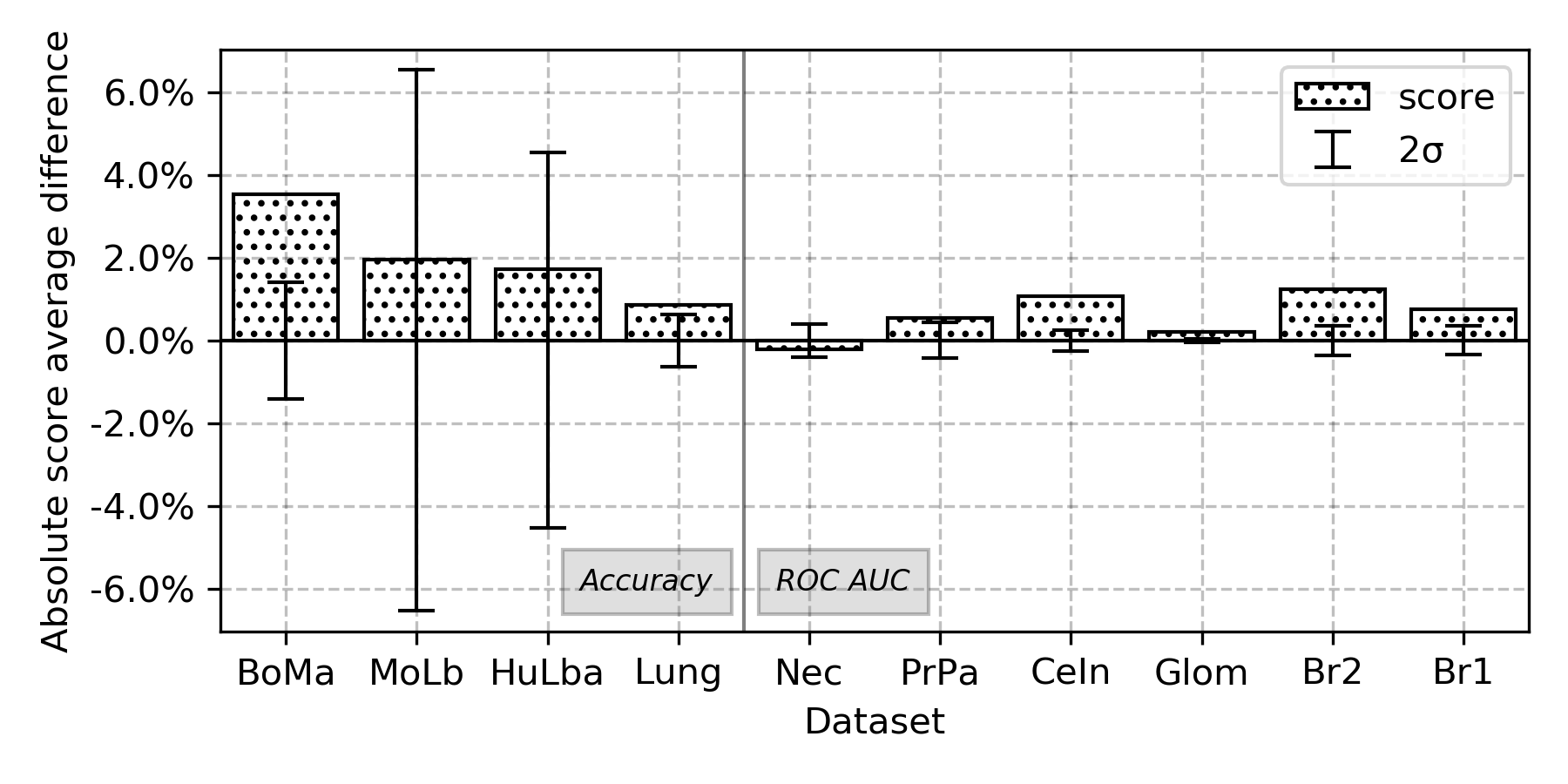}}
    \subfigure[ResNet50]{\includegraphics[scale=0.5]{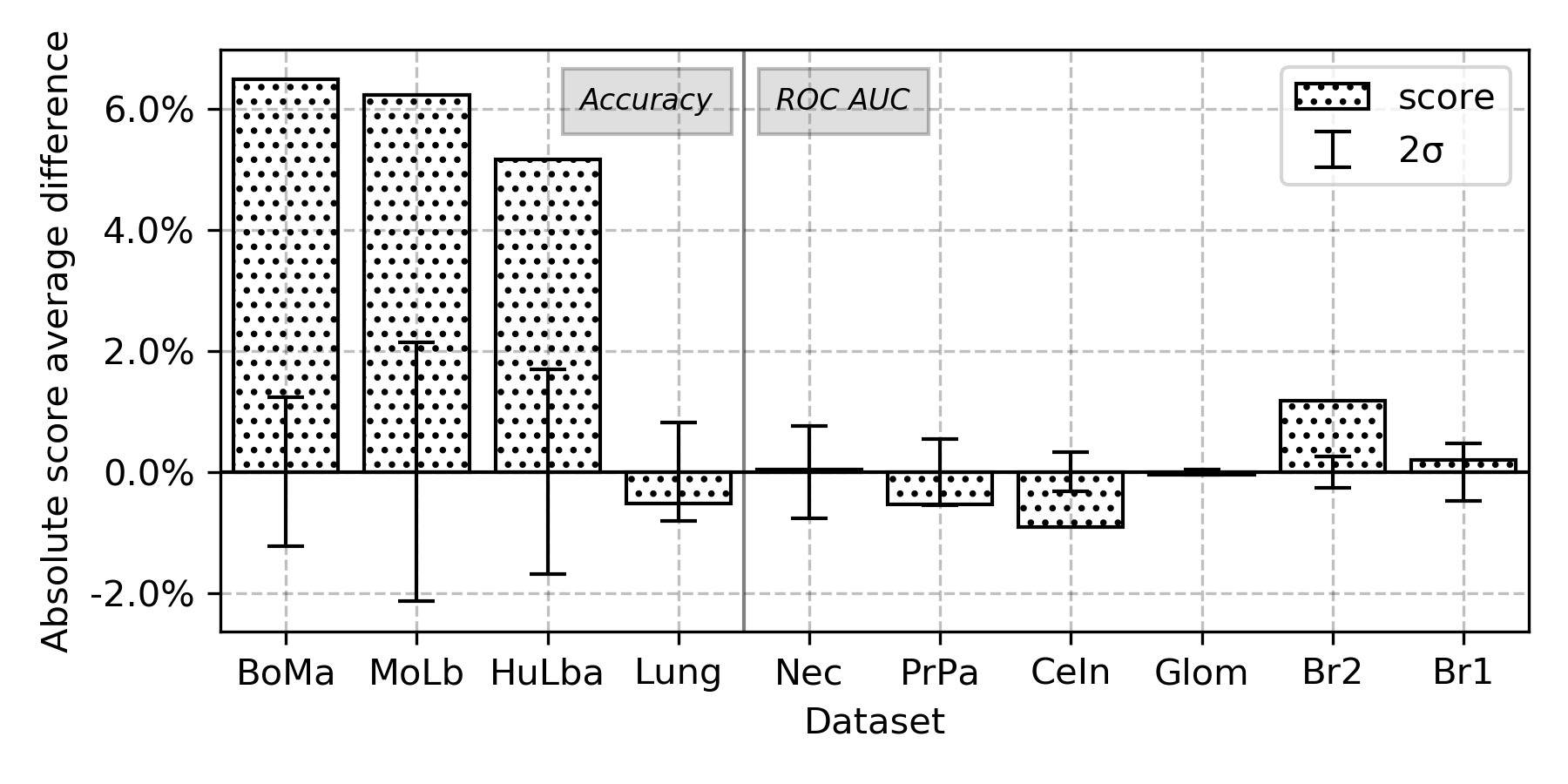}}
    \caption{Absolute score difference between multi-task versus ImageNet pre-training using \textbf{feature extraction} as transfer protocol on our ten evaluation tasks. Positive difference indicates that multi-task pre-training yield superior performance. Tasks are sorted by evaluation metric and increasing dataset size. The variability of the multi-task transfer is measured using \textit{two} standard deviations given by the error bars.}  
    \label{fig:res_featext}
\end{figure}

\begin{figure}[t]
    \centering
    \subfigure[DenseNet121]{\includegraphics[scale=0.5]{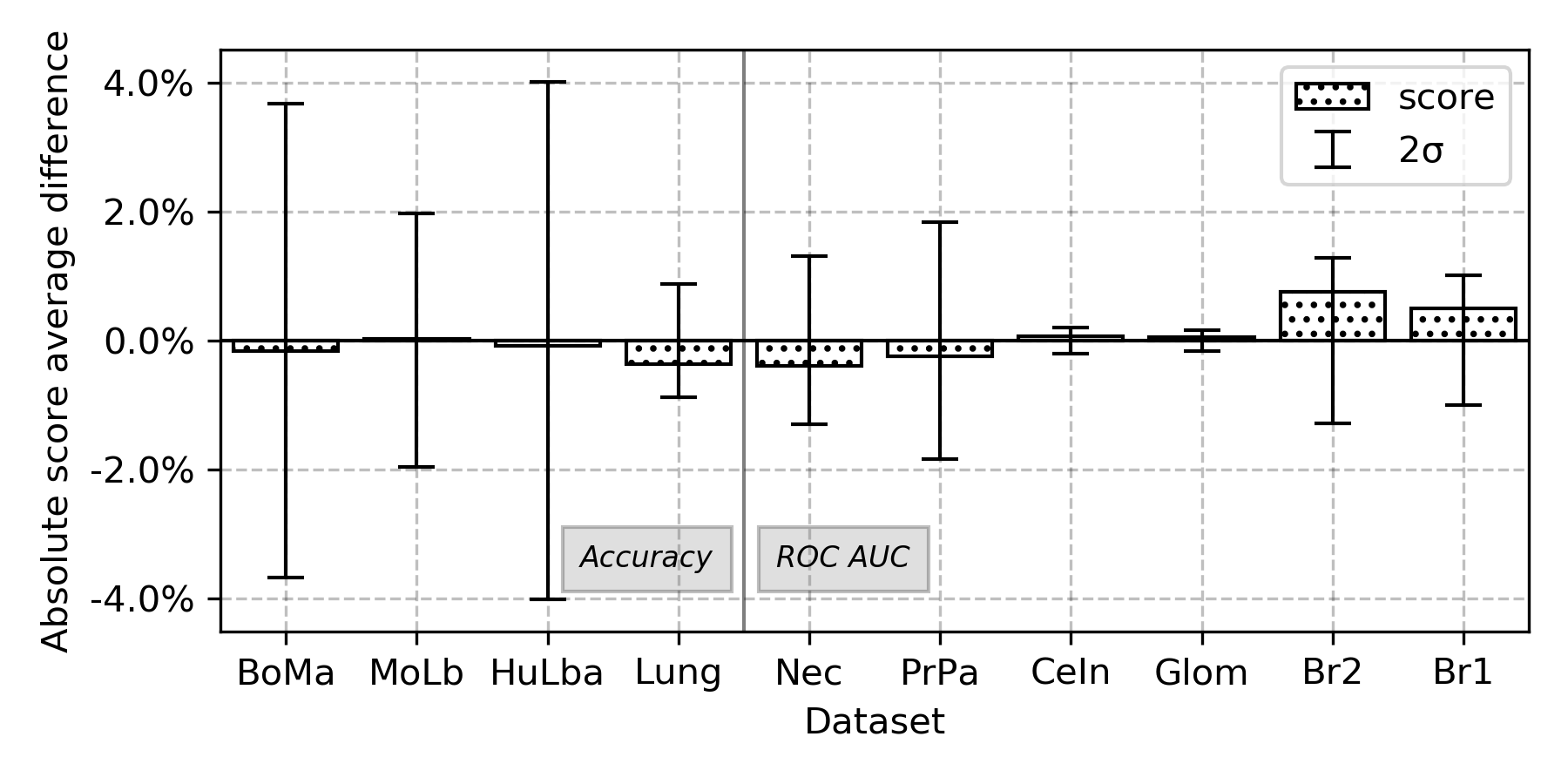}}
    \subfigure[ResNet50]{\includegraphics[scale=0.5]{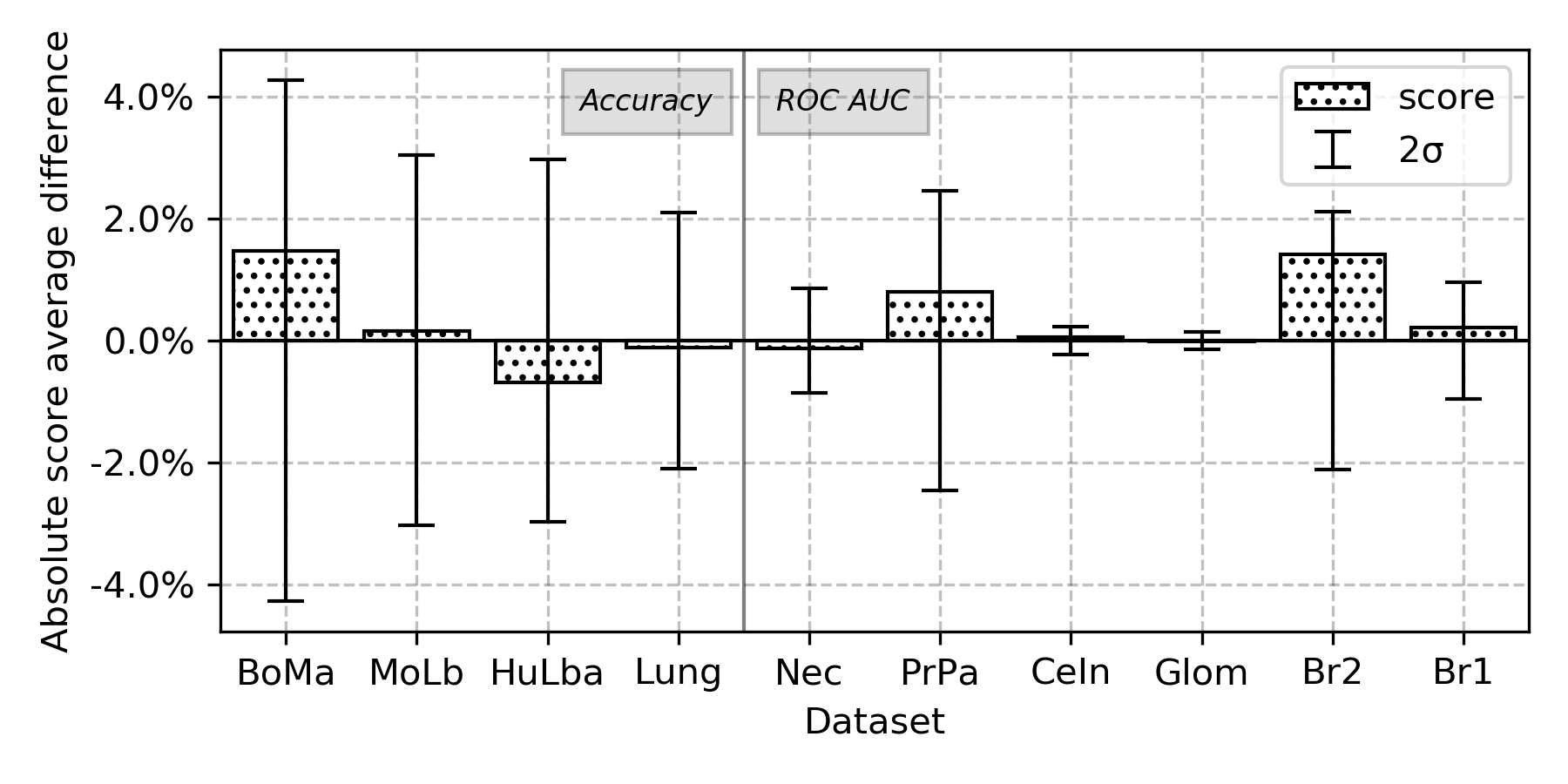}}\\
    \caption{Absolute score difference between multi-task versus ImageNet pre-training using \textbf{fine-tuning} as transfer protocol on our ten evaluation tasks. See Figure \ref{fig:res_featext} for details. Error bars are computed using two times the largest standard deviation among the ones resulting from ImageNet and multi-task fine tuning.}  
    \label{fig:res_finetune}
\end{figure}

\begin{figure*}
    \centering
    \includegraphics[scale=0.6]{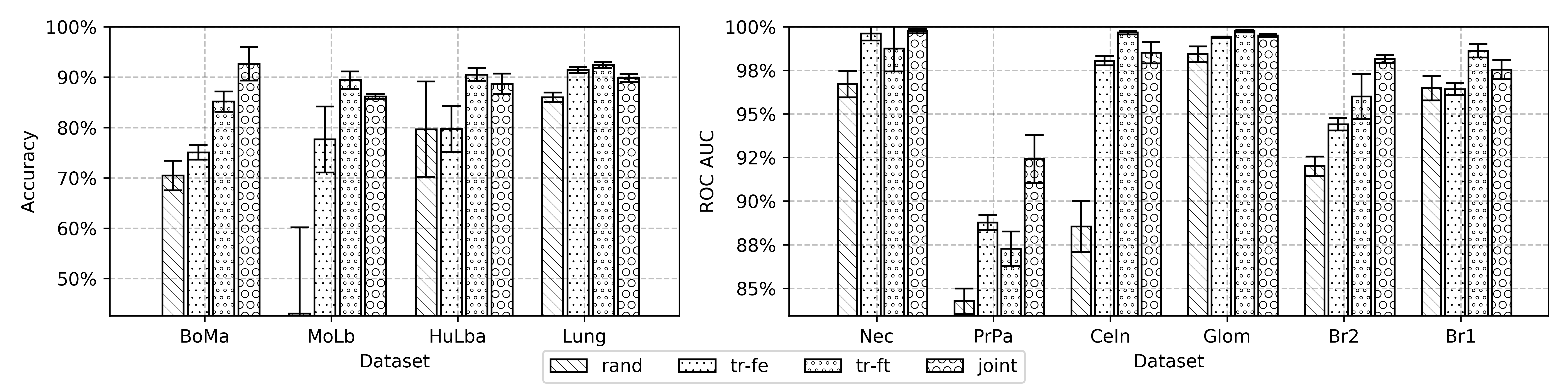}
    \caption{Performance comparison of different approaches using DenseNet121: training from scratch (\textit{rand}), feature extraction (\textit{tr-fe}) and fine-tuning (\textit{tr-ft}) using our multi-task pre-trained networks and joint training (\textit{joint}). Each bar is the average performance and its error bar gives twice the standard deviation of a given method over five runs. For exact scores, see Supplementary Table V.}  
    \label{fig:res_all_densenet}
\end{figure*}

We report how our methods compare to transfer from ImageNet, training from scratch and joint training in Section \ref{ssec:transfer_perfromance}. Then, we study the effect on transfer of the various evaluated multi-task training hyperparameters in Section \ref{ssec:hyperparameters}. Finally, we discuss our results and future works in Section \ref{ssec:res:discussion}.

\subsection{Transfer performance}
\label{ssec:transfer_perfromance}

\begin{table}[ht]
    \centering
    \caption{Evaluation tasks, their evaluation metric, training set size and full name.}
    \begin{tabular}{|c|c|c|c|}
        \hline
        \multirow{2}{*}{\textbf{Task}} & \textbf{Train} & \multirow{2}{*}{\textbf{Full name}} & \multirow{2}{*}{\textbf{Metric}}\\
        & \textbf{size} & & \\  
        \hline
        BoMa & 652 & BoneMarrow & \multirow{4}{*}{\rotatebox[origin=c]{75}{Accuracy}}\\
        MoLb & 2438 & MouseLba & \\
        HuLba & 4397 & HumanLba & \\
        Lung & 5443 & Lung & \\
        \hdashline
        Nec & 791 & Necrosis & \multirow{6}{*}{\rotatebox[origin=c]{75}{ROC AUC}} \\
        PrPa & 1346 & ProliferativePattern & \\
        CeIn & 1816 & CellInclusion & \\
        Glom & 14605 & Glomeruli & \\
        Br2 & 14953 & Breast1 & \\
        Br1 & 18261 & Breast2 & \\
        \hline
    \end{tabular}
    \label{tab:dataset_train_info}
\end{table}

As explained in Section \ref{ssec:exp:transfer_eval}, we have used both feature extraction and fine-tuning to evaluate transfer performance. We could not repeat our experiments with several datasets splits given the computational cost, which would have allowed us to perform formal statistical tests for method comparison. To ease the discussions below, we will nevertheless call significant any difference between two average errors that exceed, in absolute value, twice the maximum of their standard deviations. If the scores were Gaussian distributed, this would ensure that each average score is outside a 95\%-confidence interval around the other score.

Absolute score differences between feature extraction from ImageNet and multi-task pre-trained networks can be found in Figures \ref{fig:res_featext}a and \ref{fig:res_featext}b for DenseNet121 and ResNet50 respectively. Our DenseNet121 features yield superior scores for nine datasets out of ten (all but \textit{Necrose}) of which superiority is significant for all but two (\textit{HumanLba} and \textit{MouseLba}). The score difference is in favor of ImageNet features on \textit{Necrose} although it is not significant. ResNet50 features yield superior results for six out of ten datasets (all but \textit{CellInclusion}, \textit{Glomeruli}, \textit{ProliferativePattern} and \textit{Lung}) of which superiority is significant for all but two datasets (\textit{Necrose} and \textit{Breast1}). Out of the four datasets where ImageNet features yield superior scores, the difference is significant for only one of them (\textit{CellInclusion}). It is interesting to note that the largest difference of scores in favor of ImageNet is only $0.21\%$ (ROC AUC) on \textit{Necrose} for DenseNet121 whereas the difference in favor of our features is at most $3.52\%$ (ACC) on \textit{BoneMarrow}. Similary with ResNet50, the largest differences are $0.91\%$ (ROC AUC) on \textit{CellInclusion} and $6.48\%$ (ACC) on \textit{BoneMarrow} respectively in favor of ImageNet features and ours. Therefore, it appears that the loss of performance when our features underperform compared to ImageNet is lower than the expected gain of performance when our features are better. This indicates that the performance gain or loss you would obtain with multi-task features are dataset dependent and is hard to predict apriori, although the loss of performance is usually small compared to the possible improvement. Another interesting observation is the stability of transfer performance as only four evaluations (out of 20) exhibit standard deviations larger than 0.5\% (ROC AUC or ACC). 

Regarding fine-tuning, our features outperform ImageNet ones for five and six datasets with DenseNet121 and ResNet50 respectively (see Figures \ref{fig:res_finetune}a and \ref{fig:res_finetune}b). However, none of the differences are significant whether or not the advantage is in favor of our approach.

As a summary, feature extraction transfer approach seems to benefit from multi-task pre-training as 15 evaluations (out of 20) are in favor of our approach of which 11 are significant. Only two evaluations are significantly in favor of ImageNet features.  However, fine-tuning from our features yield comparable performance with ImageNet features initialization as no score difference is significant (11 evaluations are in favor of our approach).

As an additional experiment, we compare transfer by feature extraction and fine-tuning with a similar model trained from scratch and the joint MTL approach described in Section \ref{ssec:exp:parameters} (see Figure \ref{fig:res_all_densenet}). These experiments show that fine-tuning improves over feature extraction significantly on most datasets and that training from scratch is subpar compared with the transfer approaches, feature extraction included. Both observations confirm previously published results \cite{mormont2018comparison, ponzio2019dealing, tajbakhsh2016convolutional,shin2016deep}. It appears that joint training significantly improves the performance on small datasets (\textit{BoneMarrow} and \textit{ProliferativePattern}) over all other approaches. For larger datasets, performance seem to lie between fine-tuning and feature extraction, or to be on par with fine-tuning on the datasets where the task is almost solved (ROC AUC or ACC close to 1). 

\subsection{Study of multi-task training hyperparameters}
\label{ssec:hyperparameters}

\begin{figure}[t]
    \centering
    \includegraphics[scale=0.565]{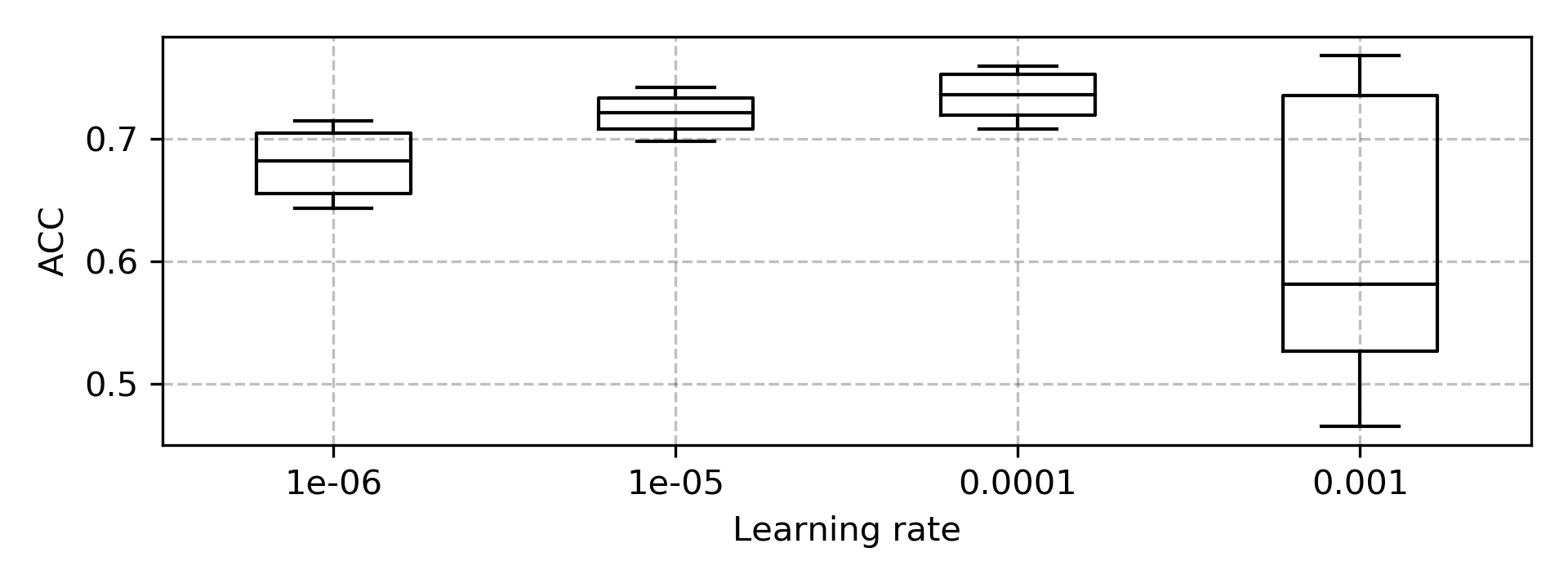}
    \caption{Distributions of scores per learning rate on DenseNet121 with HumanLba dataset. Each boxplot results from the aggregation of the transfer scores of all models using the same learning rate value on the given network and dataset.}
    \label{fig:lr_effect}
\end{figure}

\begin{figure}[t]
    \centering
    \subfigure[HumanLba]{\includegraphics[scale=0.5]{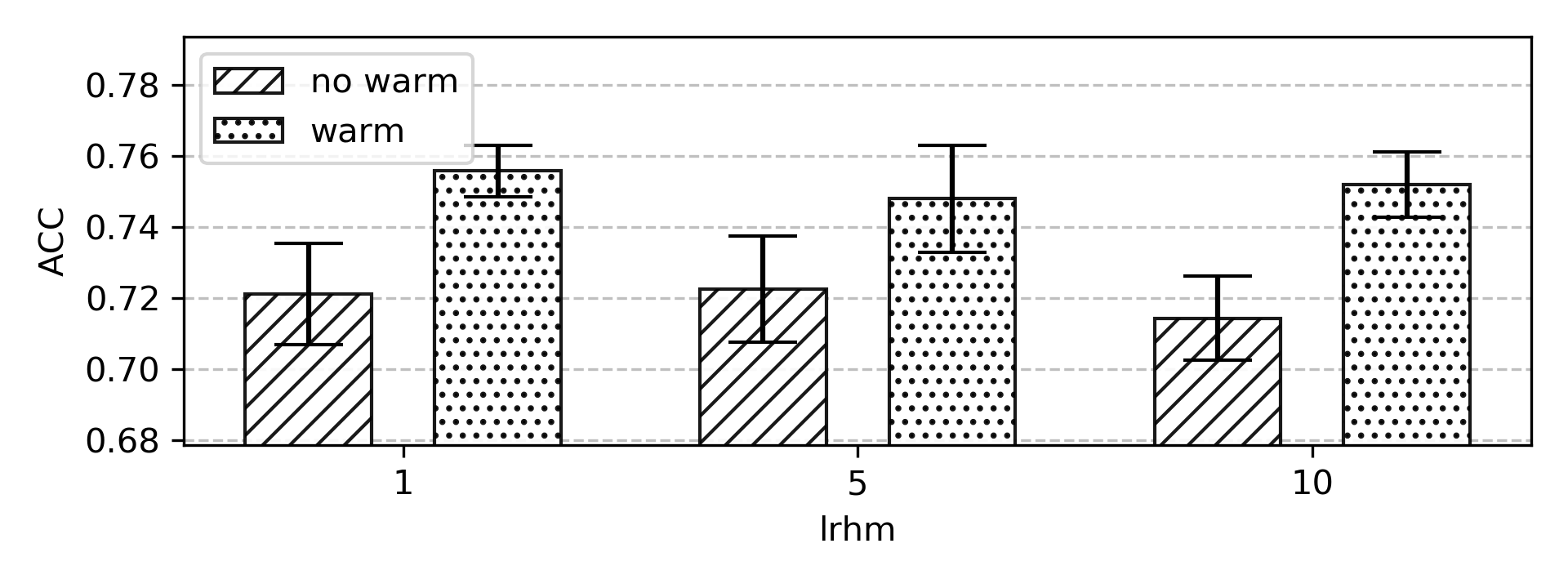}}
    \subfigure[MouseLba]{\includegraphics[scale=0.5]{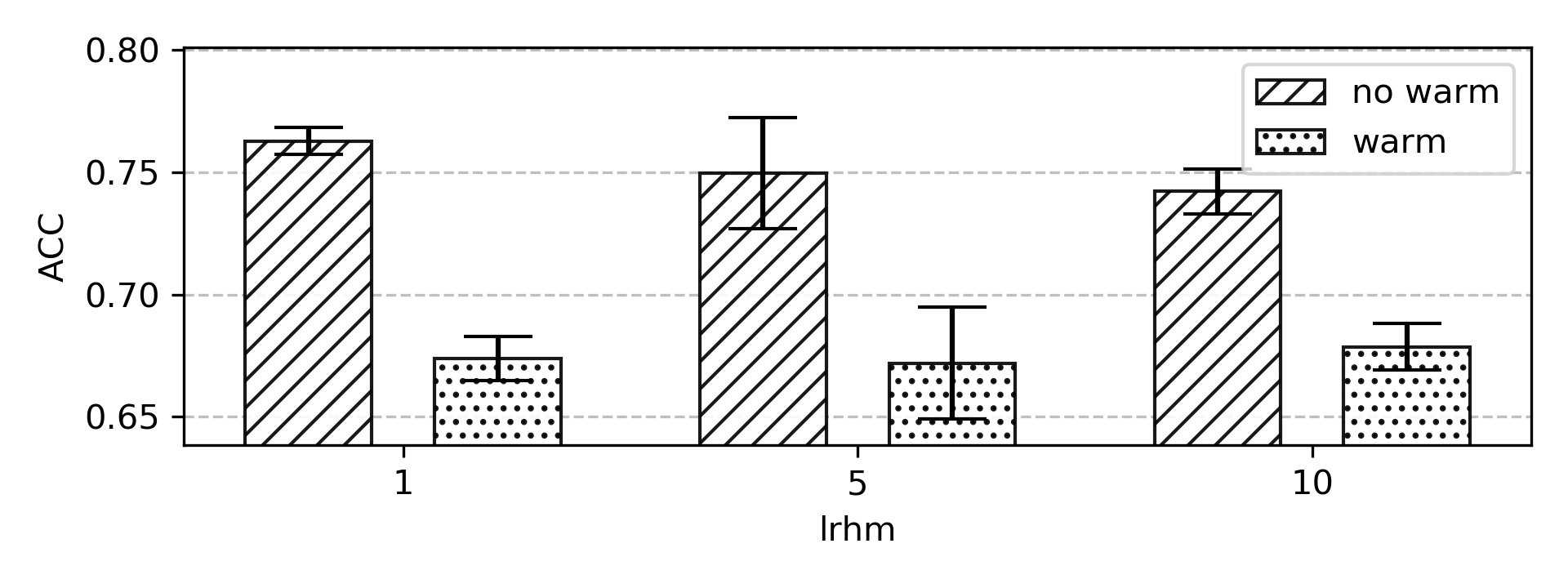}}
    \caption{Transfer performance for combinations of the hyperparameters $\gamma_\tau$ (learning rate heads multiplier, \textit{lrhm}) and $w$ (warm up) on two different datasets with learning rate $\gamma = 10^{-4}$ on DenseNet121. Error bars report twice the standard deviation.}
    \label{fig:hyper_other_effect}
\end{figure}

Our experiments have shown that the most impactful parameters on transfer performance was the multi-task training learning rate. Figure \ref{fig:lr_effect} shows the distributions of scores per learning rate using DenseNet121 and \textit{HumanLba} dataset. We have picked this plot specifically because it exhibits the most frequent pattern regarding the effect of the learning rate. Similar plots for other datasets as well as ResNet50 can be found in Section D in the Supplementary Materials. We have observed that the highest learning rate $10^{-3}$ yields highly variable performance and is most of the times inferior to lower learning rates. It indicates that this specific value is too high to cope with the multi-task setting as it prevents the models from making use of the tasks information efficiently. This is likely due to training instabilities (convergence issues, noisy training loss) and overfitting. It appears that the lowest learning rate $10^{-6}$, although yielding more stable performance, generally underperforms higher learning rates $10^{-5}$ and $10^{-4}$. For both networks and most datasets, the latter learning rate $10^{-4}$ is the best performing on average.

The impact of the two other hyperparameters $\tau_\gamma$ and $w$ seems to be minor for most datasets as variation stays within two standard deviations. Moreover, there is no pattern emerging from our experiments regarding those hyperparameters in general. Two exceptions to this observation are the \textit{HumanLba} and \textit{MouseLba} datasets which exhibit significant performance variations on both networks. The variations are shown in Figure \ref{fig:hyper_other_effect} (similar figures can be found for other networks and datasets in Supplementary Section D). Those Figures show that \textit{HumanLba} benefits from warming up whereas \textit{MouseLba} performance are hurt. This indicates that the effect on transfer performance of warming up and multiplying heads learning rate is very dataset dependent and no general rule can be drawn from our experiments.

\subsection{Discussion and future works}
\label{ssec:res:discussion}

Features extracted from our models are in general superior to ImageNet ones 
which shows that multi-task pre-training is effective at creating task-specific features. This important observation confirms the conclusions of previous works that domain-specific pre-training is a good idea and also validates the multi-task approach when a large source dataset is not available. 

The second important observation is that fine-tuning does not benefit from using our models as we obtain similar transfer performance whatever the source. This indicates that fine-tuning is able to recover the lack of specificity of ImageNet features compared to ours. It contradicts our initial hypothesis that transfer should work better when source and target domains are similar. There might be several reasons why we observe such phenomenon. First, Yosinski \etal \cite{yosinski2014transferable}, who concluded about the importance of task similarity for efficient transfer, performed their experiments on simple architectures (\eg simple stack of convolutional layers). In our case, we have used more recent architectures (\ie residual and densely connected networks) which are easier to train and might be less impacted by their initial weights. Second, most previous works have shown that fine-tuning was beneficial in a single source task transfer scenario exclusively. Our multi-task pre-training is able to learn specific features but we can not exclude that a more advanced approach could result in even stronger features that might change our conclusion that fine-tuning does not benefit from MTL transfer. For instance, some papers have highlighted training difficulties associated with multi-task (\eg gradients interference \cite{yu2020gradient}) and transfer learning (\eg batch normalization \cite{chang2019domain}) that could be investigated in our context. In addition, it is likely that, given a target task, not all tasks in the pool contribute equally to transfer performance. Some tasks might even have a destructive effect during the pre-training phase (\ie if they were removed, transfer performance would increase). Incorporating a training mechanism that could dynamically increase (\textit{resp.} decrease) the contribution of constructive (\textit{resp.} destructive) tasks would certainly help improving the resulting features. Alternatively, instead of using all  the tasks, one could find a mechanism for selecting a subset of (the most relevant) source tasks for a given target task. Such solution would entail however a significant additional computational cost, since a new MTL model would have to be trained for each new target task.

There are also several interesting research directions regarding the architecture. On the one hand, it would be interesting to study the effect of increasing the capacity of the task-specific parts. On the other hand, we have only worked with classification tasks so far but it is possible to incorporate directly segmentation or detection tasks to the pre-training by appending an ad-hoc network as a head. Hopefully, enriching training signal with a dense prediction task such as segmentation could improve the transferability of the resulting models. However, this approach also raises pratical questions and issues such as model memory usage or loss aggregation. 

More practically, we plan to integrate newly released datasets into our pool to reinforce its versatility and hopefully the transferability of the pre-trained models. We also plan to integrate the pre-trained models 
to the Cytomine open-source tool \cite{maree2016collaborative} and the BIAFLOWS benchmarking platform \cite{rubens2019biaflows}.

\section{Conclusion}

In this work, we have investigated the use of multi-task learning for pre-training neural networks in digital pathology. We have first created a pool of classification tasks from existing sources containing almost 900k digital pathology images. Using this pool, we have pre-trained a neural network in a multi-task setting in order to transfer the resulting model to unseen digital pathology tasks. Using a robust evaluation protocol, we have shown that transferring a model pre-trained in multi-task can be beneficial for the performance on the target task. When compared to transfer from ImageNet, our pre-training approach coupled with feature extraction yields comparable or better performance depending on the target dataset. We have observed that fine-tuning multi-task or ImageNet pre-trained models yields comparable performance. It suggests that fine-tuning is able to recover from the lack of feature specificity whatever the pre-training source. However, pre-training remains crucial, as models trained from scratch are clearly inferior. 

\appendices

\section*{Acknowledgments}
The authors would like to thank all previous authors and scientists who released their datasets (see Supplementary Section F) as well as Joeri Hermans and Ulysse Rubens for technical support. RaM was supported by IDEES grant with the help of the Wallonia and the European Regional Development Fund (ERDF). Computational infrastructure is partially supported by ULi\`ege, Wallonia, and Belspo funds. 

\ifCLASSOPTIONcaptionsoff
  \newpage
\fi

\bibliography{jbhi2019}

\end{document}